\documentclass{ieeeaccess}
\usepackage{cite}
\usepackage{amsmath,amssymb,amsfonts}
\usepackage{algorithmic}
\usepackage{graphicx}
\usepackage[utf8]{inputenc}
\usepackage[T1]{fontenc}
\usepackage{textcomp}

\usepackage{bm}
\makeatletter
\AtBeginDocument{\DeclareMathVersion{bold}
\SetSymbolFont{operators}{bold}{T1}{times}{b}{n}
\SetSymbolFont{NewLetters}{bold}{T1}{times}{b}{it}
\SetMathAlphabet{\mathrm}{bold}{T1}{times}{b}{n}
\SetMathAlphabet{\mathit}{bold}{T1}{times}{b}{it}
\SetMathAlphabet{\mathbf}{bold}{T1}{times}{b}{n}
\SetMathAlphabet{\mathtt}{bold}{OT1}{pcr}{b}{n}
\SetSymbolFont{symbols}{bold}{OMS}{cmsy}{b}{n}
\renewcommand\boldmath{\@nomath\boldmath\mathversion{bold}}}
\makeatother

\def\BibTeX{{\rm B\kern-.05em{\sc i\kern-.025em b}\kern-.08em
    T\kern-.1667em\lower.7ex\hbox{E}\kern-.125emX}}

\begin{document}
\history{}
\doi{}

\title{Is Open Source the Future of AI? A Data-Driven Approach.}

\author{\MakeUppercase{Domen Vake}\authorrefmark{1,2},
\MakeUppercase{Bogdan Šinik}\authorrefmark{1},
\MakeUppercase{Jernej Vičič}\authorrefmark{1}, and
\MakeUppercase{Aleksandar Tošić}\authorrefmark{1,2}}

\address[1]{DIST, UP FAMNIT, Glagoljaška 8, 6000 Koper, Slovenia 
(e-mail: domen.vake@famnit.upr.si; bogdan.sinik@famnit.upr.si; jernej.vicic@upr.si)}
\address[2]{IP, InnoRenew CoE, Livade 6a, 6310 Izola, Slovenia 
(e-mail: aleksandar.tosic@upr.si)}

\tfootnote{}

\markboth
{Vake \headeretal: Is Open Source the Future of AI? A Data-Driven Approach.}
{Vake \headeretal: Is Open Source the Future of AI? A Data-Driven Approach.}

\corresp{}

\begin{abstract}
Large language Models(LLMs) have taken center stage in both academia and industry. Alongside discussions on their usability, accuracy, and societal impacts are growing concerns regarding privacy, transparency, and their potential misuse in illicit activities. A critical topic within these debates is the trustworthiness of LLMs, specifically when models in question are proprietary. A frequently proposed solution to significantly improve trustworthiness is open-sourcing models. However, this option has significant drawbacks such as illicit applications where models can be modified and misused, lack of incentive structures (mostly financial) to support open-sourcing, and protecting intellectual property. On the other hand, most LLMs are trained by the private sector due to data and computing resource requirements. These costs are significant and proprietary models are better positioned to achieve a return on investment. 
There are also other approaches that lie somewhere on the spectrum between completely open-source and proprietary. These can largely be categorised into open-source usage limitations protected by licensing, partially open-source (open weights) models, hybrid approaches where obsolete model versions are open-sourced, while competitive versions with market value remain proprietary.

Currently, discussions on where on the spectrum future models should fall on remains unbacked and mostly opinionated where industry leaders are weighing in on the discussion. In this paper, we present a data-driven approach by compiling data on open-source development of LLMs, and their contributions in terms of improvements, modifications, and methods. Our goal is to avoid supporting either extreme but rather present data that will support future discussions both by industry experts as well as policy makers.

Our findings show that the open-source community can significantly contribute to improving the performance of open-source models, assuming that performance metrics are representative of realistic model performance. We identified positive patterns in open-source contributions, such as a significant reduction in model size with a manageable decrease in accuracy. Moreover, we identify patterns in community growth and the architectures that benefit disproportionately from its engagement.

\end{abstract}


\titlepgskip=-21pt

\maketitle

\section{Introduction}
\label{sec:introduction}
\PARstart{A}{rtifcial} intelligence, particularly large language models (LLMs), is an important topic in the computer industry at present. Despite the numerous fears and dogmas surrounding AI, it is certain that it has become integral to our life. This research specifically concentrates on a particular subset of the field, known as open-source AI. As the name implies, it is a form of Artificial Intelligence that is accessible to all individuals. However, the term open-source encompasses a range of approaches, from fully open models with unrestricted access to their code and weights, to partially open models that share certain components (e.g., weights or training datasets) and impose restrictions through licenses. There is considerable debate on whether this type of technology should be universally accessible. Our aim was to investigate whether the open-source community is actively contributing to the field, regardless of differing philosophical convictions. Due to their substantial computational requirements, running LLMs on personal computers was previously impractical. However, the development of increasingly compact versions with impressive capabilities is leading to a significant transformation. It is now feasible to run your own model, provided it is sufficiently small, on a home computer's graphics processing unit (GPU), even if the GPU is several years old \cite{patel2023google}.

In the field of LLMs, the open-source community relies heavily on the willingness of large corporations to release their models in the open source domain. This transition is understandably hindered by challenges from competitive markets, as well as the high barrier to entry, given that developing such models requires a significant investment in both time and money \cite{patel2023google, jiang2023empirical}.

The future direction of the industry is uncertain. On one hand, keeping models closed-source (CSM) allows companies to maintain their competitive edge, which could significantly impact revenue; on the other hand, OSMs provide much needed transparency, fostering greater competition, which could lead to a more dynamic and progressive market. Concerns with OSMs in relation to competition are further amplified due to the ease with which high-performing models can be used to train other models (assuming similar architecture).

OpenAI has occasionally released open-source models, such as GPT-2, but has shifted towards keeping its more advanced models, like GPT-4, proprietary. This strategy suggests a cautious approach, balancing open innovation with concerns about misuse. For instance, Meta has taken an open-source approach by releasing models like Llama2 and Llama3, while models like Gemini and Anthropic's Claude remain closed source.
Proponents of closed source argue that unrestricted availability of OSMs could pose significant risks if misused, enabling the creation of tools that are potentially harmful to society. However, arguments in support of OSMs highlight the importance of transparency and fostering innovation to allow for broader scrutiny of model behavior. Additionally, the incremental improvement of open models would allow for participation in the incremental development of guardrails for these models.
Regardless of where readers fall on the spectrum, there is value in supporting these opinions and ideas with data driven analysis. 

With the release of multiple OSMs, the need for objective metrics to evaluate their performance has become increasingly important. In response, the open-source community has developed various metrics to assess performance across different task domains, such as mathematics, logic, and medicine. 
Hugging Face \footnote{https://huggingface.co/} has emerged as the primary platform where the open-source community shares contributions, including models, datasets, and methods.  It also serves as a hub for benchmarking models and facilitating research. 

Due to its widespread use, Hugging Face provides a comprehensive view on the entire OS community and is considered a primary source of information for this study. The aim of our research is to analyse the open-source environment in the field of LLMs in an effort to provide industry experts, stakeholders, and policy makers with data to support informed decision-making and to help guide the future direction of AI in general. Accordingly, we identified the following research questions:

\begin{enumerate}
    \item Does the open-source community influence the development of LLM models?
    \item Is it possible to quantify this impact in terms of performance?
    \item Can we observe patterns and trends for possible future directions?
\end{enumerate}

\section{Literature review}
Due to its continuous growth, Hugging Face has emerged as the leading platform for sharing machine learning (ML) and artificial intelligence (AI) models, resulting in increasing levels of complexity. A relational database called HFCommunity was established to facilitate the analysis and resolution of this issue \cite{HFCommunity}.

Castaño et al. \cite{castano2024analyzing} conducted research with a similar goal, focusing on temporal changes on Hugging Face. They examined various trends, including fluctuations in user count, models, commits, and overall platform activities. Additionally, they analysed model maintenance, and categorised them into two groups. They used all Hugging Face data accessible via the HF Hub, in addition to the HFCommunity database established by \cite{HFCommunity}. However, their study overlooked the influence of Hugging Face on the AI community.

The article by Patel et al. \cite{patel2023google} highlights the significance of the open-source AI community and explains its rapid growth in the wake of major industry leaders like Google, Microsoft, and OpenAI. A significant milestone in this area was the release of the LLama model, and the open-source community promptly recognised the possibilities and potential involved in this release.

As previously mentioned, open-source AI models offer an extensive range of possibilities. At a recent conference, the authors \cite{CombiningModels} demonstrated the effective use of Hugging Face models. Given the significant challenges in developing a model with broad intelligence, the researchers integrated ChatGPT capabilities with models from Hugging Face using an agentic architecture.  This approach yielded impressive results in multiple domains. ChatGPT was tasked with creating a plan of action and assigning specific duties to each open-source model based on their own areas of expertise. This is an excellent demonstration of the influence and capabilities of the open-source community.

Fine-tuning has become a critical strategy for adapting large language models (LLMs) to specific tasks while leveraging the extensive knowledge embedded in pre-trained models. Numerous studies on open-source language models have proposed a parameter-efficient approach to fine-tuning, which greatly reduces the computational resources required when adapting the model to a specific task. Fine-tuning is especially prominent in the Hugging Face ecosystem, where models can be adapted for diverse applications with minimal overhead.\cite{han2024parameter}

In addition to fine-tuning, merging models has gained attention as a technique for integrating the capabilities of multiple models for broader applicability. Wang et al. \cite{yang2024model} systematically explored strategies for effectively merging large language models, highlighting the potential for combining complementary models to address complex tasks. Hugging Face serves as a central hub for this practice, providing tools to combine checkpoints and enable cross-model functionality.

The article \cite{vulnerbilities} examines the security risks associated with open-source AI. A much higher number of repositories with high vulnerabilities were found compared to those with low vulnerabilities, particularly in root repositories. This emphasises the importance of ensuring the security of the technology in order to facilitate its utilisation.

In a recent paper \cite{pepe2024hugging}, authors analysed the transparency of Hugging Face's pre-trained models regarding database usage and licenses. The analysis revealed that there is often a lack of transparency regarding the training datasets, inherent biases, and licensing details in pre-trained models. Additionally, this research identified numerous potential licensing conflicts involving client projects. Of the 159,132 models analysed, merely 14 percent of these models clearly identified their datasets with specific tags. A detailed examination of a statistically significant sample comprising 389 of the most frequently downloaded models showed that 61 percent documented their training data in some form.

\section{Methodology}
To address our research questions, we first needed to collect data. Given our focus on impact, general data from repositories was not especially beneficial. Instead, we collected data from the Open LLM Leaderboard on Hugging Face \cite{open-llm-leaderboard}, where we obtained information on repositories of models that are currently on the leaderboard and models that are awaiting evaluation for the leaderboard through scraping. A Python pipeline was developed to clean and enrich the available data on GitHub\footnote{https://github.com/VakeDomen/HF\_analysis}. The leaderboard data includes model architecture and precision as well as the model type and performance on the following benchmarks: ARC\cite{clark2018think}, HellaSwag\cite{zellers2019hellaswag}, MMLU\cite{hendrycks2021measuring}, TruthfulQA\cite{lin2022truthfulqa}, Winograde\cite{DBLP:journals/corr/abs-1907-10641} and GSM8K\cite{DBLP:journals/corr/abs-2110-14168}. In addition to the data provided on the leaderboard, we received supplementary information about the given models using the HF API client. This included details on repository contributors, tags, base models, used datasets, and repository activity. It is important to note that HuggingFace does not enforce the reporting of this information. Due to the self-reported and optional nature of the data, many models lack this information. The leaderboard also includes duplicates since the developers can replace models in the repository with different models under the same name. As a result, these duplicates share identical repository data but have distinct performances.  Since it is not possible to programmatically determine the current model within the repository, we selected the best-performing model to represent the repository when removing duplicates. Thus, all datasets were prepared for further use. Additionally, MistralAI has pushed updates to weights to the Mistral model naming versions v0.1 and v0.2, respectively. For a period of time the repositories referencing the base Mistral model did not specify the version used. For the specified time we sifted through the models by hand to determine the base model. The following analysis was conducted using the R programming language, focusing primarily on obvious trends. The data was categorised using several criteria, such as model type, model architecture, and parameter count. The data was initially selected and aggregated to ensure that all crucial components were easily accessible. Any models that were flagged were excluded from the dataset. In addition, we collected and analysed data on the authors' activities. To fill in the missing components, we extracted the information from tags. Furthermore, we manually reviewed the top contributors to the repositories and removed all identifiable automated bot accounts from the top 100 contributors list.
\section{Results}

\begin{figure}[ht]
    \centering
    \includegraphics[width=1\linewidth]{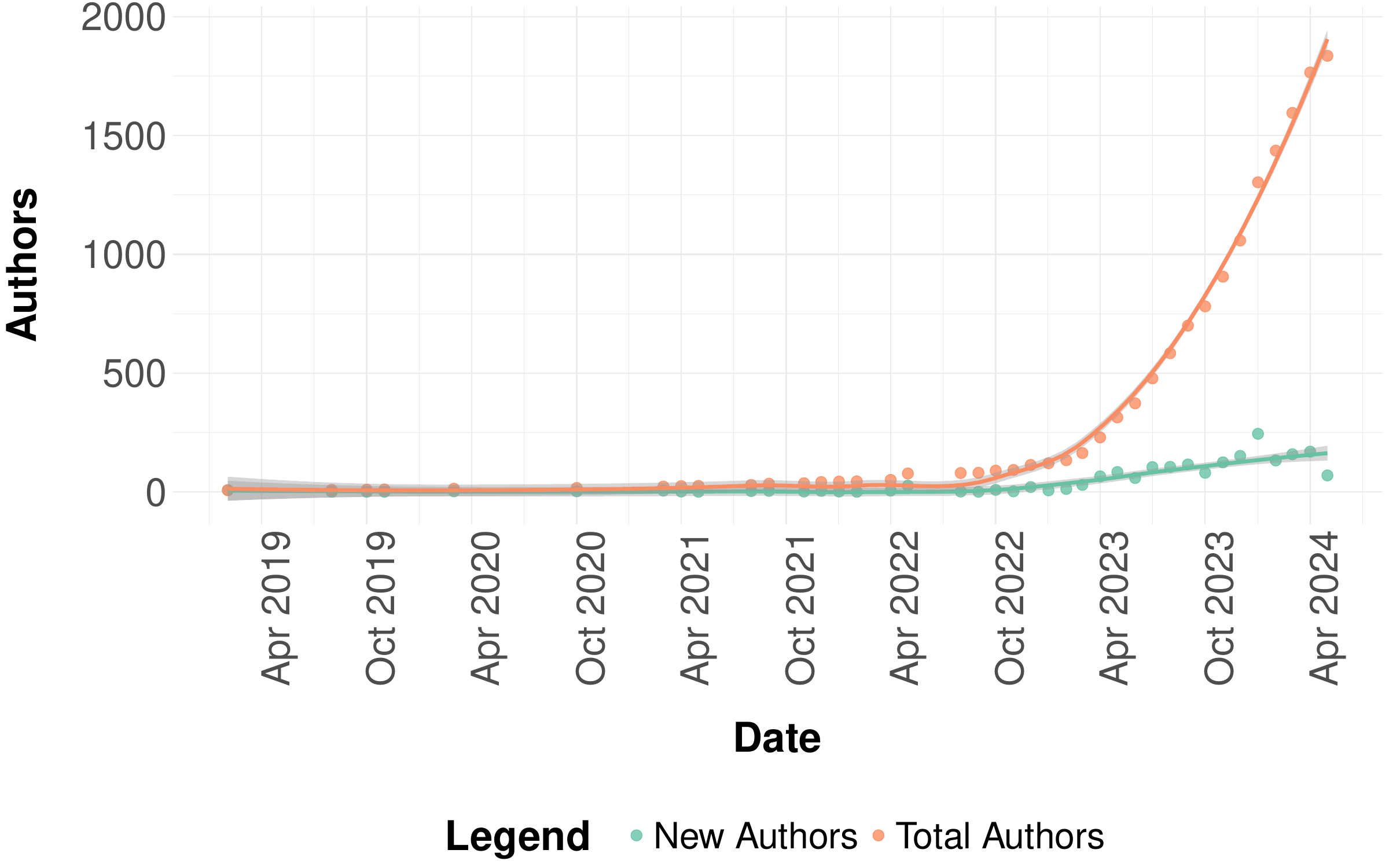}
    \caption{Total authors and new authors on the leader board over time}
    \label{fig:TotalAndNewAuthorsOverTime}
\end{figure}

The results clearly demonstrate the rapid expansion of the open-source community in Artificial Intelligence, evidenced both by the number of authors and models on the leaderboard. As for the number of authors, we cannot claim that the total number of users on Hugging Face follows the same trend, but it is clear that this area is becoming increasingly competitive as more and more distinct users appear on the leaderboard each day. As we can see in Figure \ref{fig:TotalAndNewAuthorsOverTime}, the total number of users and new authors has been rapidly increasing per day. Since mid-2023 both have shown a linear increase, indicating that LLM leaderboard is following the same trend as the Hugging Face platform, as  analysed by Castaño et al. \cite{castano2024analyzing} in their research.

\begin{figure}[ht]
    \centering
    \includegraphics[width=1\linewidth]{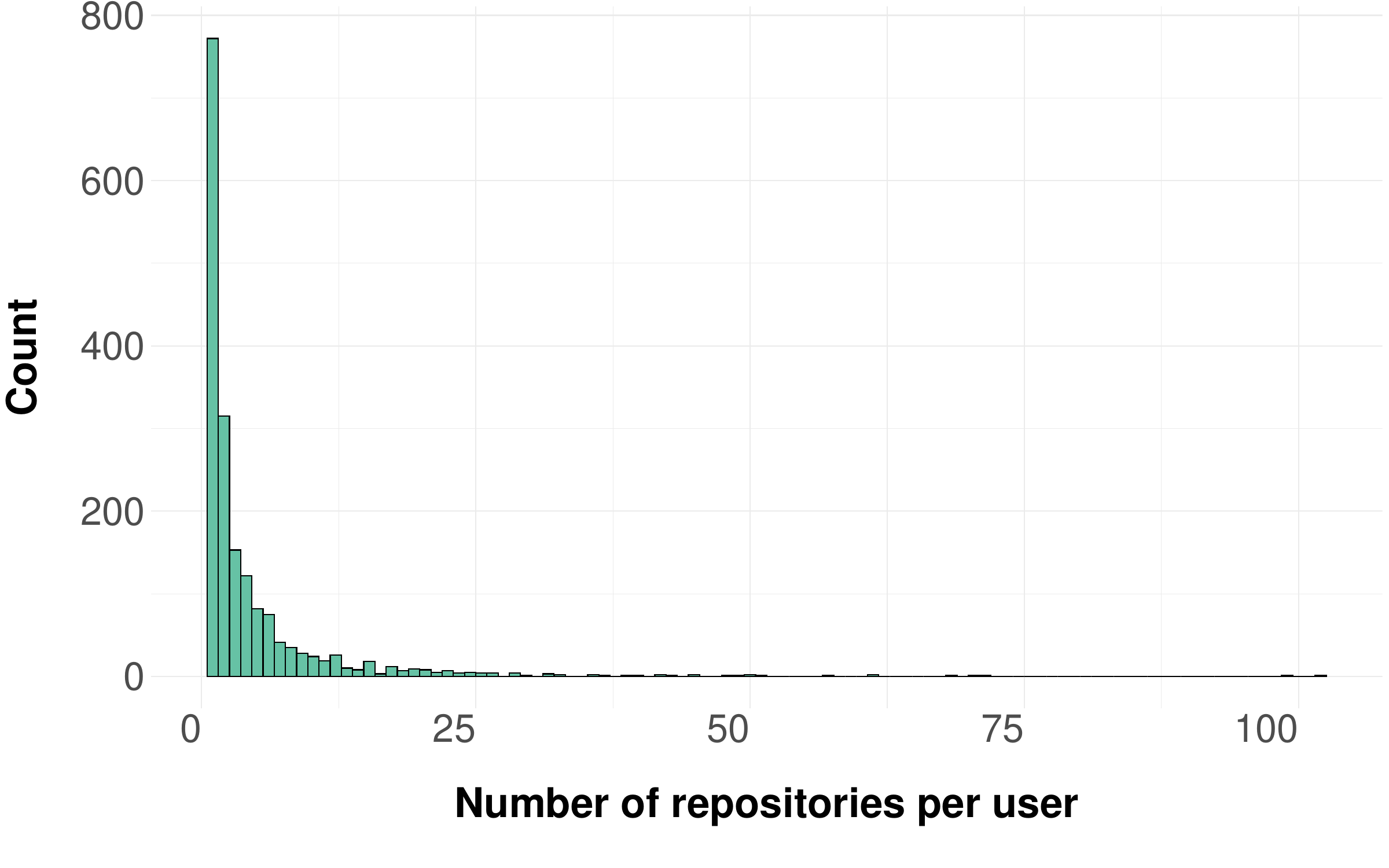}
    \caption{Distribution of number of repositories per author}
    \label{fig:repos_per_author}
\end{figure}
\begin{figure}[ht]
    \centering
    \includegraphics[width=1\linewidth]{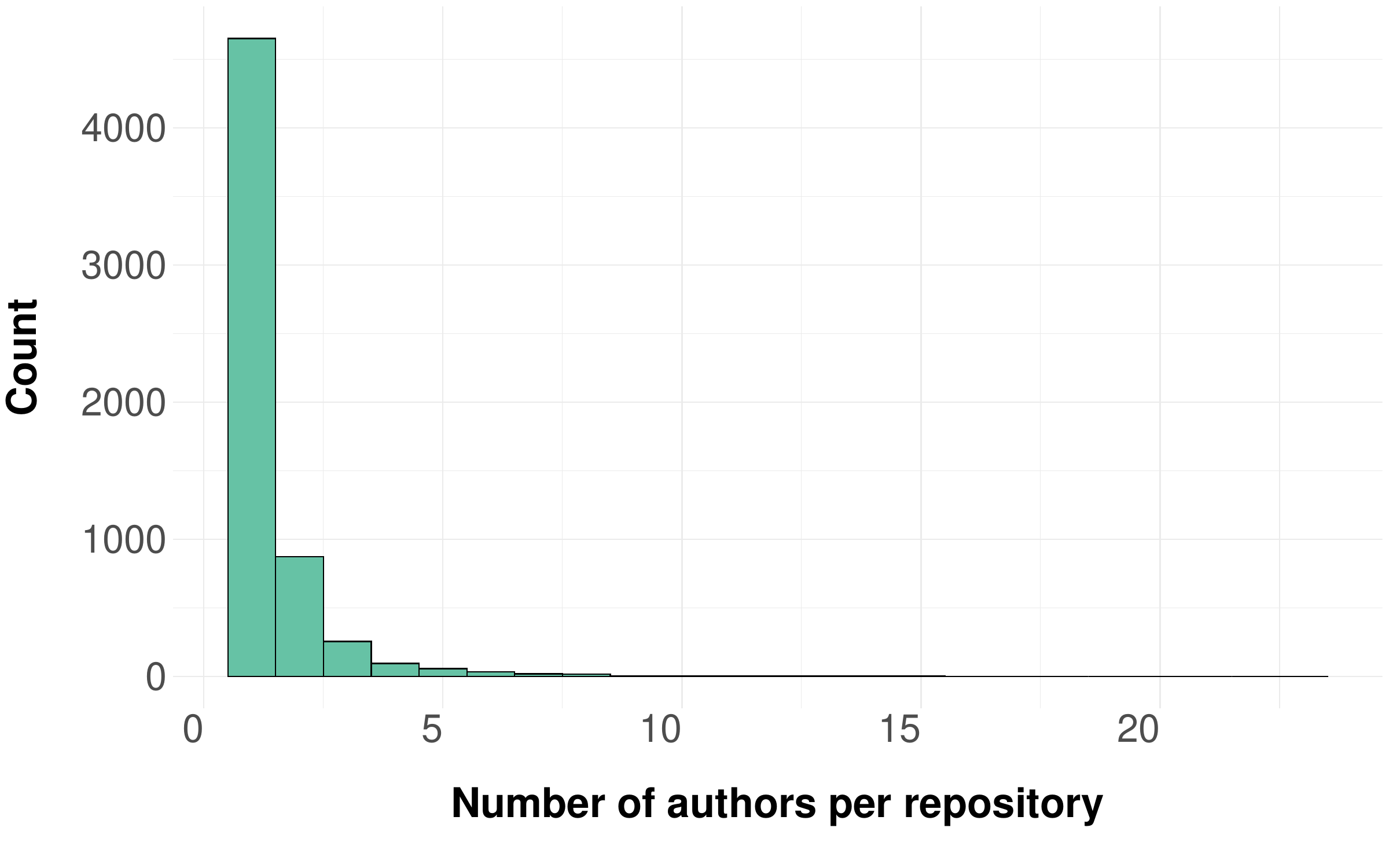}
    \caption{Distribution of number of authors per repository}
    \label{fig:authors_per_repo}
\end{figure}

The distribution of repositories among authors in Figures \ref{fig:repos_per_author} and \ref{fig:authors_per_repo} shows that the activity of authors is far from a normal distribution. The histogram shows a highly right-skewed distribution, where out of a total of 1829 authors, the top 10 users were responsible for 8\% of the total models, while the top 200 contributed to around 50\% of the total models. This concentration of users within a small circle raises concerns about the diversity and inclusivity within the community. It also reflects Price's law, which states that most of the work is typically performed by a small fraction of the workforce.

Despite the significant increase in contributors, their collaboration is at a sub-optimal level. As shown in Figure \ref{fig:authors_per_repo} a large number of repositories are created and maintained by a single profile. The distribution is highly skewed toward individual contributions, suggesting a lack of coordinated efforts that include several contributors. This pattern represents the challenges faced by collaborative networks within this community, primarily due to technological difficulties. Additionally, it is common for models to be published by smaller research teams that are frequently attributed to a single individual or profile. Even when models are the product of team effort, these teams are often closed and do not foster open collaboration, which limits the flow of ideas and expertise within the community. While publishing models on platforms like HuggingFace facilitate some level of collaboration, it often involves individuals or teams working independently rather than collectively coordinating efforts. Based on the findings, it would be beneficial to concentrate on promoting more integrated and collaborative approaches to enhance the potential and expand the boundaries of the open-source community.

\begin{figure}[ht]
    \centering
    \includegraphics[width=1\linewidth]{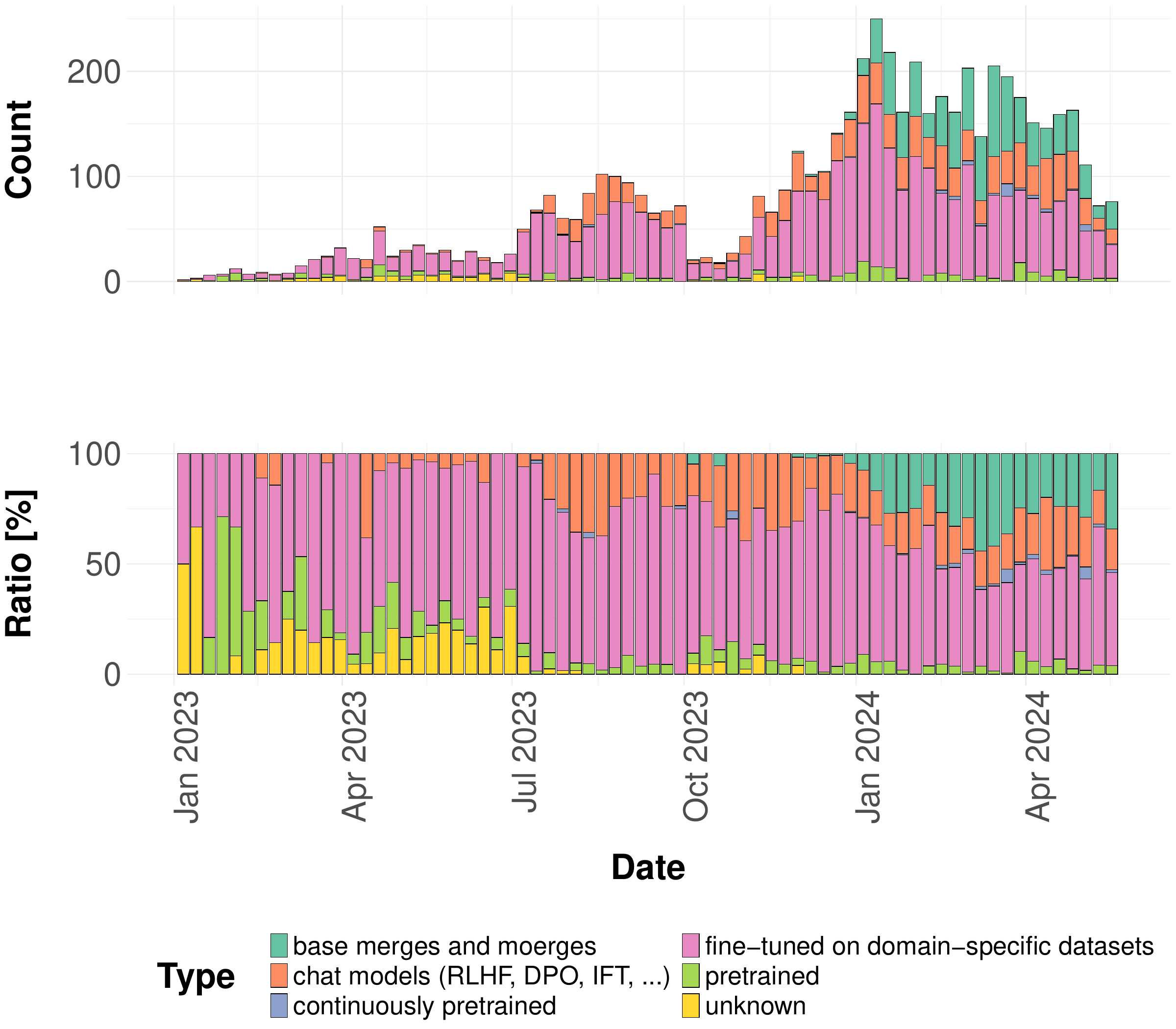}
    \caption{Number of new models per week by type}
    \label{fig:new_models_per_week}
\end{figure}

We were particularly interested in the types of models represented on the leaderboard. Hugging Face categorises models into five types: \textit{pre-trained}, \textit{continually pre-trained}, \textit{fine-tuned on domain-specific datasets}, \textit{chat models}, and \textit{base merges and merges}. Figure \ref{fig:new_models_per_week} highlights trends in model types over time. The prevailing pattern indicates a constant yet uneven increase in the number of newly released models. It is evident that there have been no instances of unknown types in recent months. This suggests that the open-source community is increasingly adopting a more rigorous and professional approach. 
The fine-tuned models and chat models seem to represent a larger proportion of the models. However, it is important to note that the dataset represents models submitted for the leaderboard benchmark testing and  not the whole HuggingFace ecosystem. This suggests that the sample includes models for which the authors believed they could compete with the best-performing models on the platform. Therefore, it can be reasonably assumed that the models in the sample are competitive in terms of performance.
Mergers, a relatively recent trend appears to be growing in popularity. This demonstrates a need within the community for more sophisticated and proficient models. As stated in the literature review, combining models can yield impressive results since each model specialises in  specific tasks. The rationale behind merging these models is to create more comprehensive models. 

\begin{figure}[ht]
    \centering
    \includegraphics[width=1\linewidth]{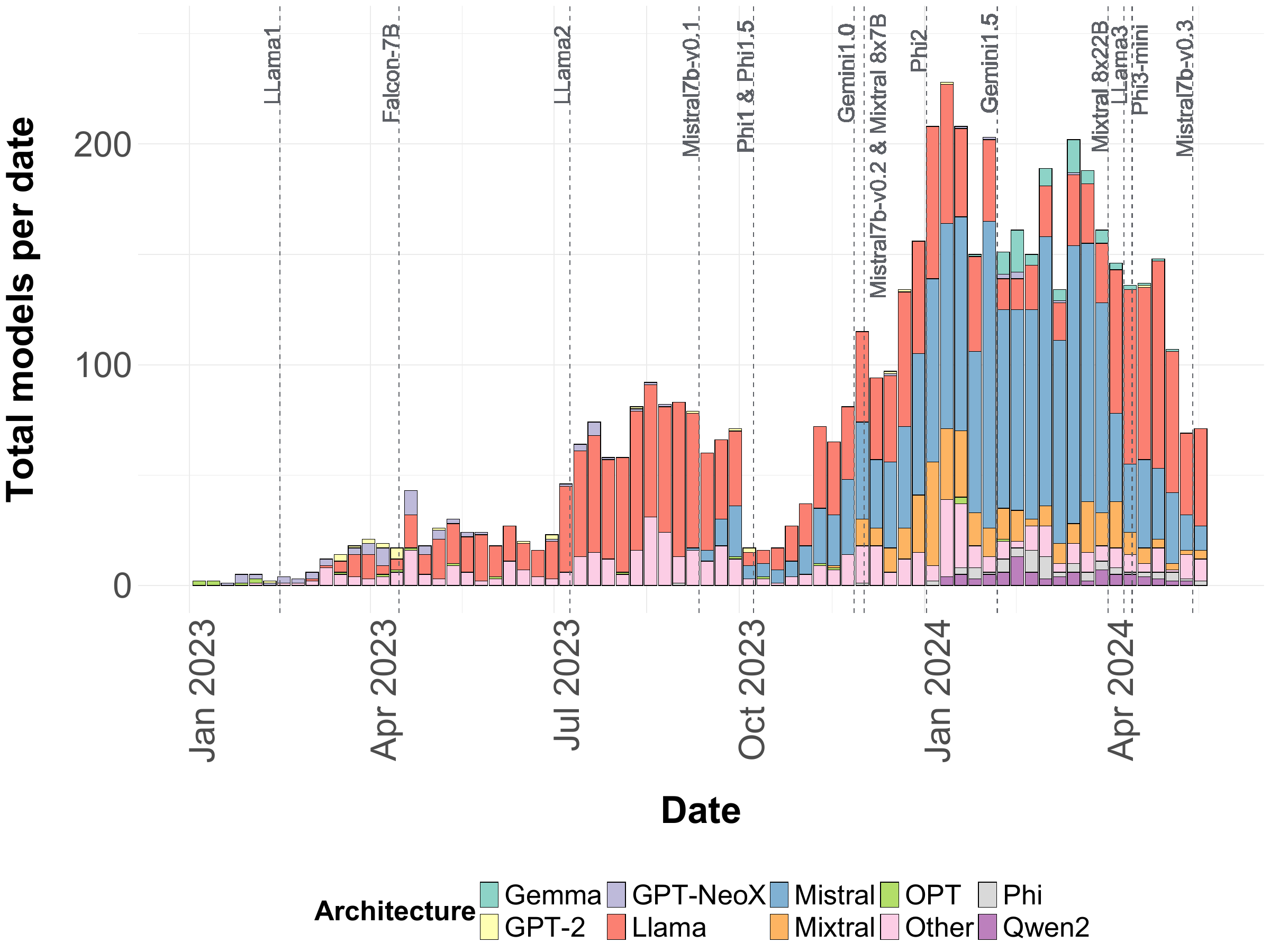}
    \caption{Total models per week based on architecture}
    \label{fig:models_per_week_by_architecture}
\end{figure}

\begin{figure}[ht]
    \centering
    \includegraphics[width=1\linewidth]{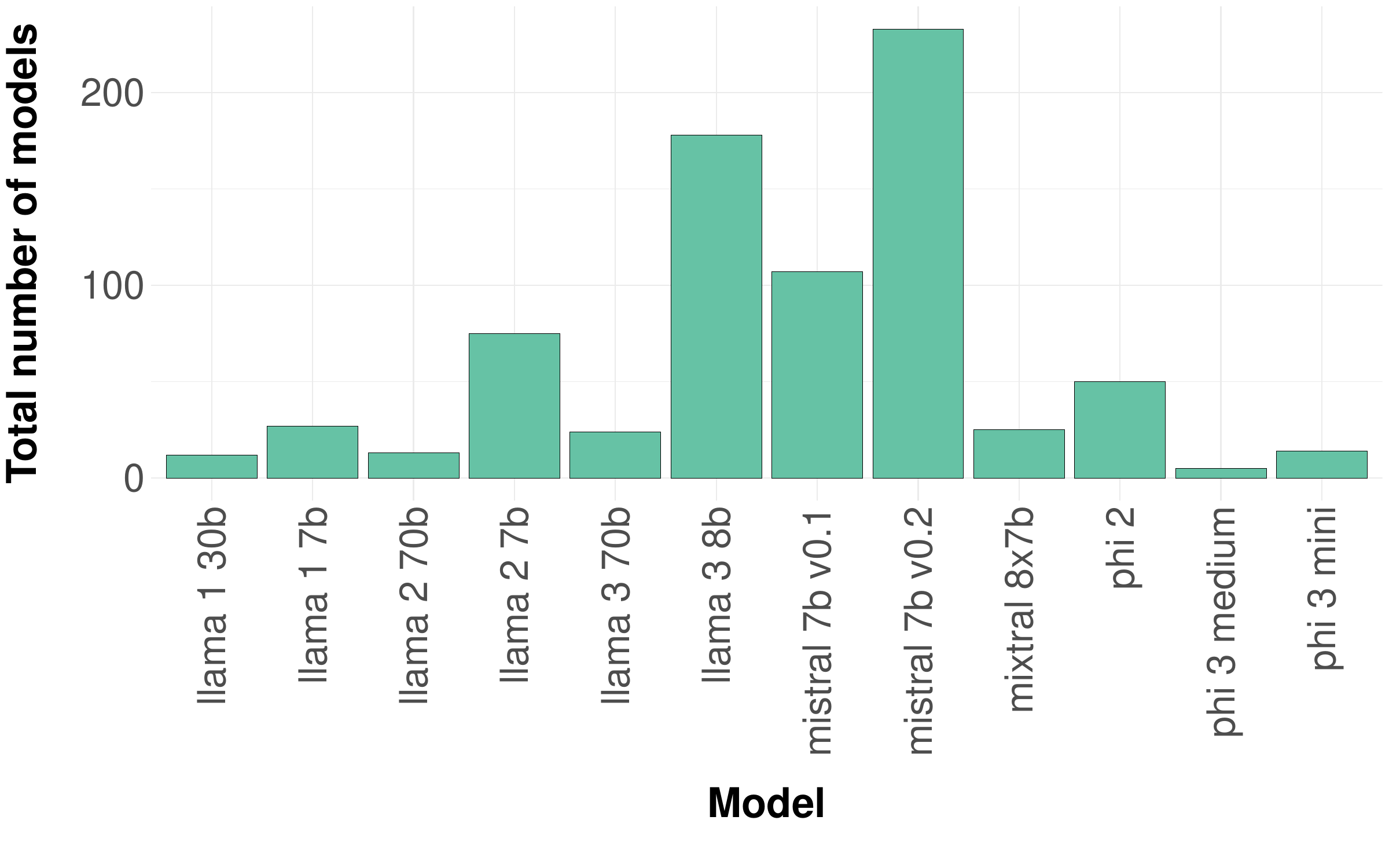}
    \caption{Total number of models for each base model}
    \label{fig:totalPerBase}
\end{figure}

In addition to analysing model types, we examined different model architectures. Figure \ref{fig:models_per_week_by_architecture} illustrates the weekly distribution of each important model architecture. Nine architectures were selected based on the frequency of occurrences: Gemma, GPT-2, Mistral, Opt, Phi, GPT-NeoX, LLaMA, Mixtral, and Qwen2. All architectures that contributed to a minimum of 5 percent of models were picked, and the remainder were categorised as 'other.' Our analysis also considered significant model releases from the respective companies, as the introduction of new models can significantly impact the popularity of their architectures. This is particularly true for the most prominent architectures such as LLama and Mistral as shown in figure \ref{fig:totalPerBase}. The leaderboards include over 300 models that specify Mistral as the base model and over 200 that specify Llama3. Interestingly, when there is a new release from a popular source, a large portion of the experimental focus switches to the newly released model. However, not all releases trigger the same shift. The popularity of the source seems to play a large part in determining which model families tend to shift the experimentation in their favor. The Llama and Mistral models produced the largest shifts in popularity, while model families like Gemma and Phi received comparatively minor attention. 

\begin{figure}[ht]
    \centering
    \includegraphics[width=1\linewidth]{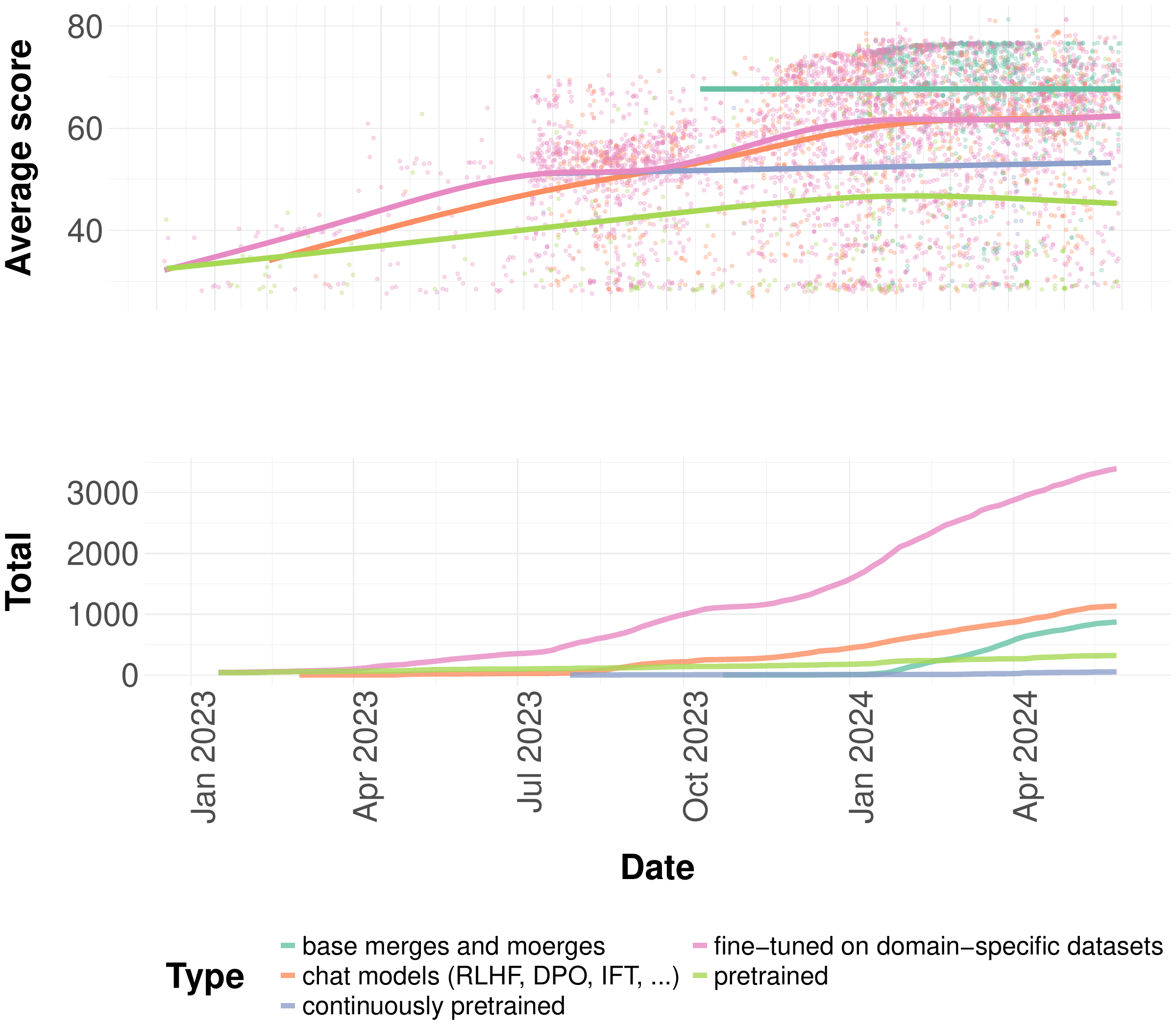}
    \caption{Average benchmark score of each model type over time paired with total number of models per type}
    \label{fig:changeOfScoreOverTime}
\end{figure}

It is important to note that an increase in the number of models on the platform does not necessarily indicate progress, but rather an increase in popularity. After examining the distribution of various model types, we attempted to assess their performance over time, based on the average score across the six benchmarks stated in the methodology. This is depicted in Figure \ref{fig:changeOfScoreOverTime}, which presents two aspects. The first aspect is the change in the average benchmark score for each model type over time. The individual dots denote the performance of a specific model, and a smooth line was added to indicate the average performance of the group that the model belongs to at a given time, making it easier to observe the temporal variations for a specific model type. It is evident that most categories improved their benchmark scores over time. Currently, the majority of the dots are clustered in the upper section of the y-axis, within an accuracy range of 60 to 80 percent. The smoothed results exhibit a similar trend. However, the one exception is the 'base mergers and mergers' category, which failed to achieve a significant rise in score. This may be due to its relatively recent introduction and its initial high position compared to others.

The visualisation in figure \ref{fig:changeOfScoreOverTime}, illustrates the cumulative quantity of models for each model type over time. Unsurprisingly, fine-tuned models constitute the majority, since this represents the most straightforward approach to model development. Pre-trained models are relatively few because developing a competitive model from scratch is extremely difficult as it requires superior data and advanced hardware resources typically available only to large corporations. Interestingly, the number of merged models is rapidly approaching that of chat models, despite their increase, beginning in 2024.

\begin{figure*}[ht]
    \centering
    \includegraphics[width=0.8\textwidth]{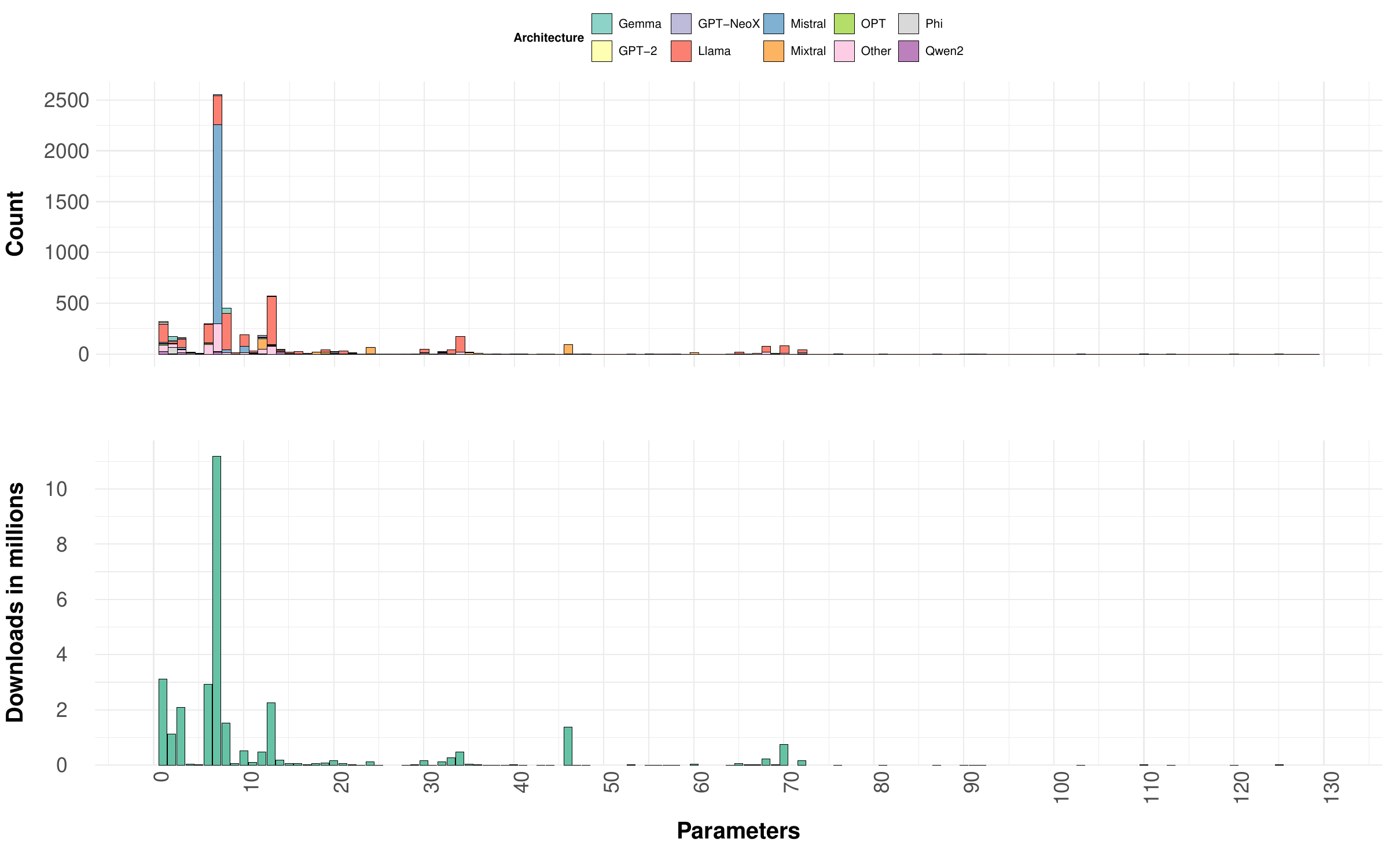}
    \caption{Distribution of models based on number of parameters colored by model architecture}
    \label{fig:paramsHistogram}
\end{figure*}


Being at the top of the leaderboard is often highlighted, but it does not tell the whole story. Models of various sizes occupy positions across the leaderboard, with many impressive achievements found in the lower ranks of the absolute benchmark scores. Therefore, the impact of model size on its popularity was examined next. Figure \ref{fig:paramsHistogram} illustrates the distribution of models based on the number of parameters, with bars colour-coded according to model architecture. The data suggests that most of the models have less than twenty billion parameters, which indicates that smaller models are favored. This pattern could be attributed to the users' ability to run the models on local machines, as well as the ease of adapting them using free resources provided by services like Google Colab\cite{googlecolab} and Kaggle\cite{kaggle}. In the second part of figure \ref{fig:paramsHistogram} we observe that the distribution between model sizes is similar to the total number of downloads. Approximately 85\% of all downloads are distributed among small models with at most 15 billion parameters. This supports the hypothesis that the open source community mainly interacts with the models that they can use in a local setting.

\begin{figure}[ht]
    \centering
    \includegraphics[width=1\linewidth]{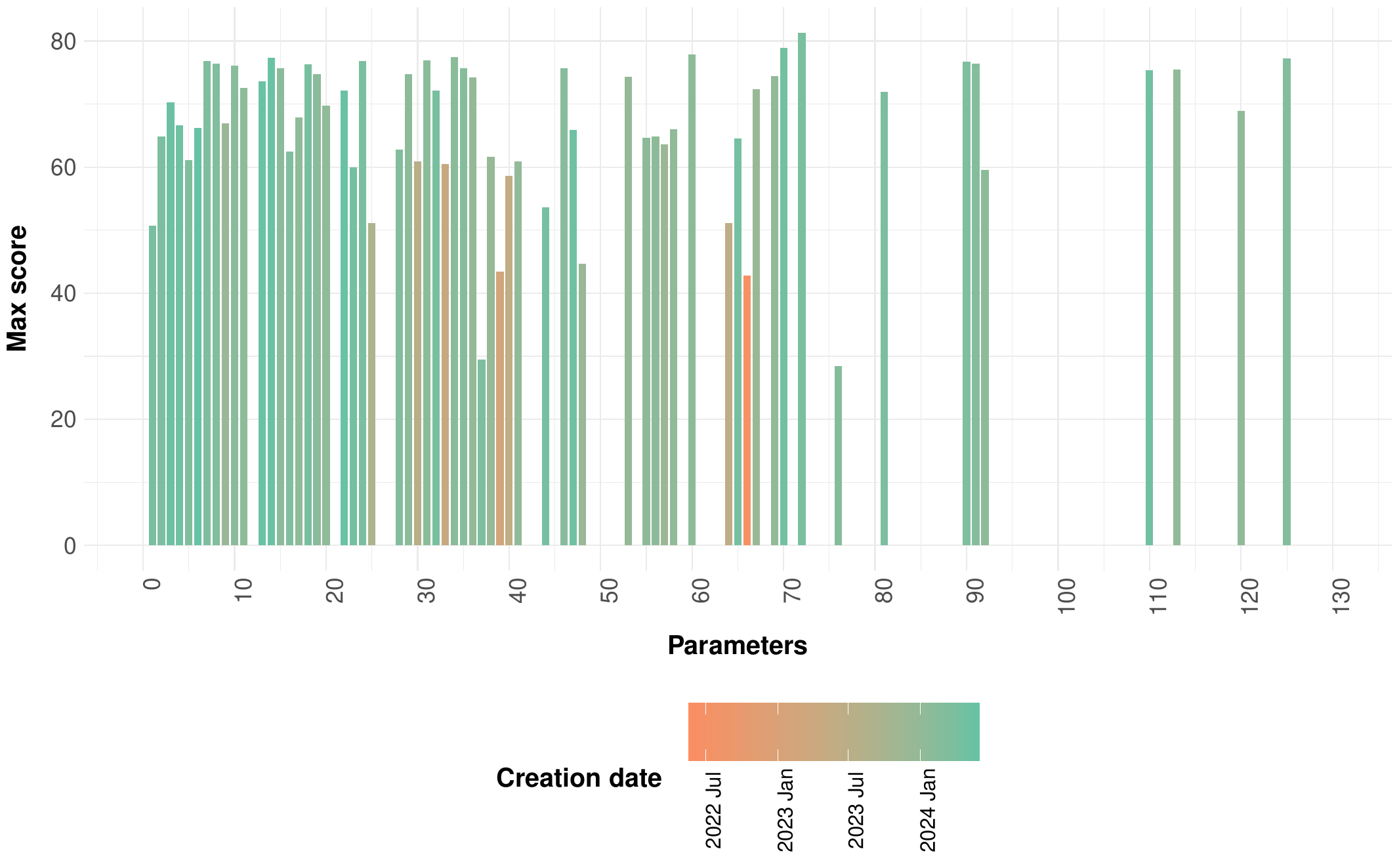}
    \caption{Model size (in billions of parameters) vs max average benchmark score}
    \label{fig:parameters_vs_max_score}
\end{figure}

An interesting observation can be made in the figure \ref{fig:parameters_vs_max_score}, which highlights the best-performing model in each size category. Larger models do not significantly outperform smaller models. Some of the highest-performing models with fewer than 20 billion parameters achieve results comparable to larger and more sophisticated models. This is particularly true for the latest models, coloured light green.


\begin{figure}[ht]
    \centering
    \includegraphics[width=1\linewidth]{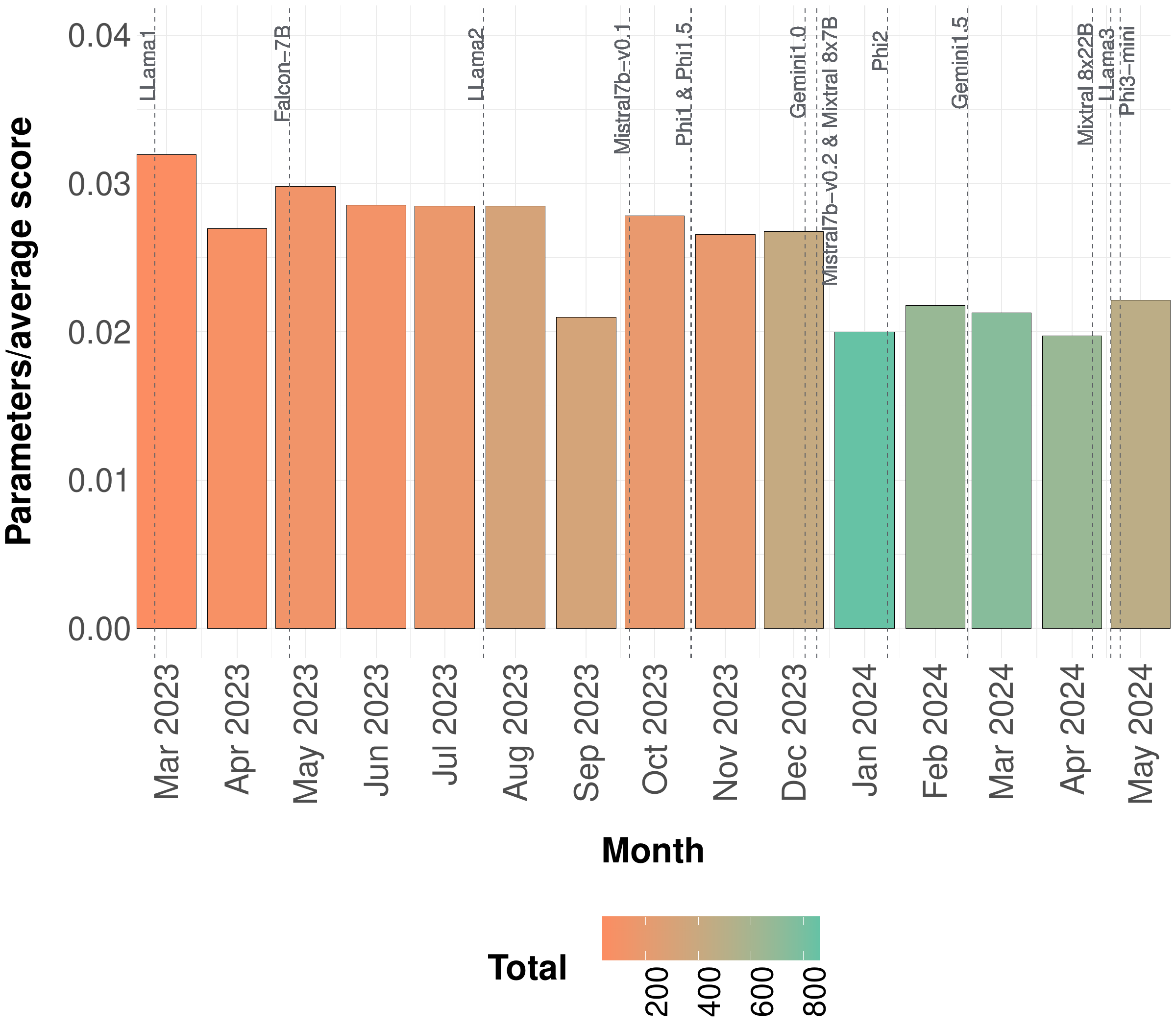}
    \caption{Average benchmark score vs number of parameters over time}
    \label{fig:avg_vs_params}
\end{figure}

Figure \ref{fig:avg_vs_params} illustrates the number of parameters needed to achieve a given performance level, displaying parameters normalised by the average benchmark score for the best model uploaded in a given month. This information was used to assess how the capabilities of the models increase with time. The color gradient in the chart represents the total number of models uploaded during each time span. The analysis shows a progressive decrease in the average size of models required to achieve the same performance, indicating that comparable or superior scores are now achievable with smaller models. While this trend reflects the community's attempt to improve the released models it is largely a consequence of well-resouced groups developing better base models that serve as a basis for further experimentation. This is encouraging for the open-source community, as high-performance models are becoming increasingly accessible to people without large amounts of computing resources.

The most challenging task was measuring Hugging Face's contributions to advancing the AI field, particularly the impact of individuals focused on enhancing existing base models rather than creating new ones from scratch. Given that base model data of fine tunes is available for most leaderboard entries, we aimed to evaluate how significantly these contributors improved the most popular models through fine-tuning and other optimisation techniques. This analysis highlights the crucial role that community contributions play in refining and advancing AI capabilities, even when starting from pre-existing foundational models. Merges were excluded from this analysis because it was not possible to attribute specific contributions to the enhancement of the model. Often, merges didn't include the base models or included only one. Additionally, some merged models were created from other merged models, with the same base model appearing in multiple merge steps, making it difficult to assess the improvement to specific base models. 

\begin{figure}[ht]
    \centering
    \includegraphics[width=1\linewidth]{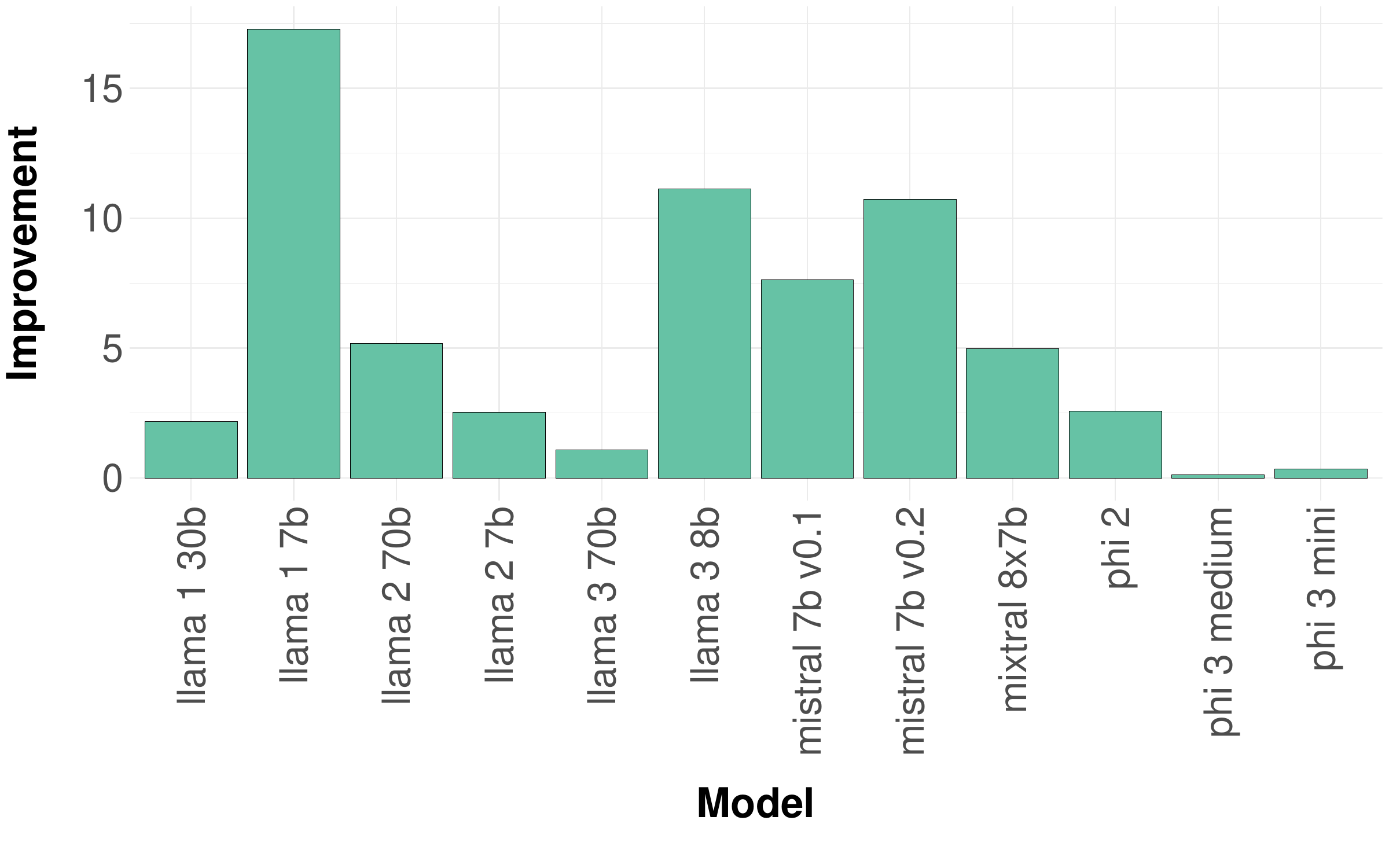}
    \caption{Total improvement of each base model}
    \label{fig:baseModelImprovements}
\end{figure}

Based on Figure \ref{fig:baseModelImprovements} the total percentage improvement of each base model was analysed. Notably, Llama 1 (7b parameters) and Mistral 7b parameters show the highest percentage improvements, demonstrating significant community engagement and successful fine-tuning efforts. In contrast, models like Phi 3 show minimal improvements, suggesting they may already be highly optimised, less susceptible, or less appealing for further fine-tuning by the community. When compared with figure \ref{fig:totalPerBase}  we can observe that the models with the highest improvements, such as Llama and Mistral, also have a substantial number of models derived from them. This correlation suggests that the community's focus on these models has led to a higher number of derivative models and to significant performance enhancements. This highlights the synergistic effect. where popular base models attract more contributors, leading to more refined and optimised versions.

Interestingly, while Mistral and Llama3 have seen significant improvements and much experimentation, Llama1 shows the most improvement despite having a relatively lower number of models built upon it. This phenomenon could be attributed to several factors. Firstly, Llama1 might have been an early model that laid the groundwork for subsequent models, making it a foundational base model with much to improve on. Secondly, the community might have identified specific areas for improvement in Llama1 that were more straightforward to address, leading to substantial enhancements with fewer derivative models. Additionally, the model is older and at that time the community was still evolving, which reflects the amount of derivative work. However, in relative terms, it still represented a significant portion of all the models uploaded at that time.

\begin{figure*}[htb]
    \centering
    \includegraphics[width=0.8\textwidth]{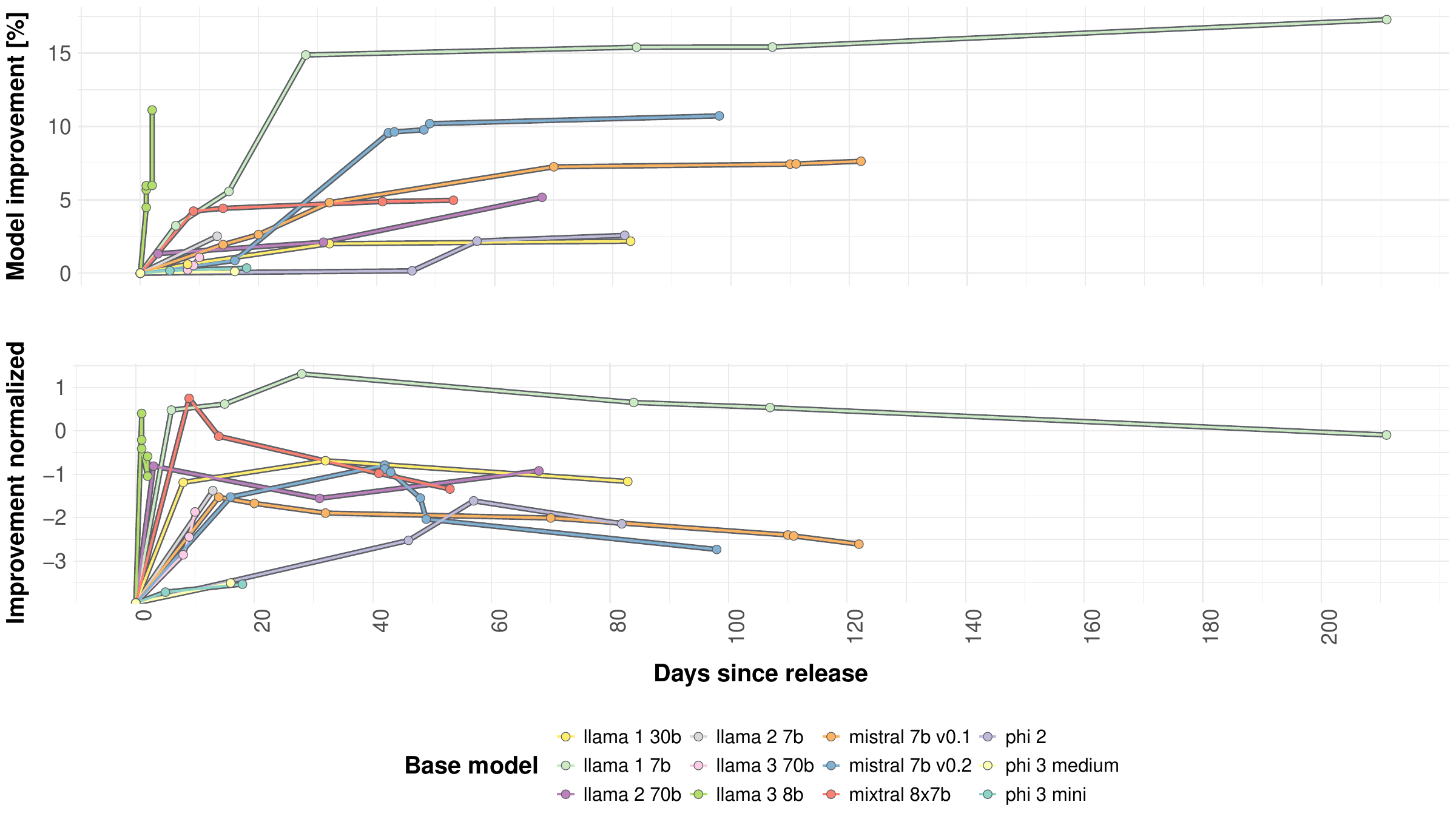}
    \caption{Evolution of most popular base models}
    \label{fig:modelEvolution}
\end{figure*}

We then extracted the models released by Meta, Microsoft, and Mistral AI and identified all incremental upgrades that used these models as their base model and were subsequently fine-tuned to achieve improved average scores on benchmarks. Additionally, we incorporated the number of days elapsed since the release of the base model, as illustrated in Figure \ref{fig:modelEvolution}. In the top section we denote all the incremental increases to the models. In the bottom section the same incremental improvements are presented, but normalised by the logarithm of the number of derivative models of the base model. This takes into account the popularity of each specific model, but also takes into account that initial improvements are easier to achieve, and the more work is done on a particular model, the harder it is to find additional improvments. This means that models that appear more trainable should rise higher on the chart. Additionally, the faster the model falls off, the more difficult it is to achieve further improvements on the model. Several observations can be made here. For example, Llama 3, featuring 8 billion parameters, experienced an enhancement above 10 percent within just a few days post-release. This was likely influenced by the immense excitement that preceded its release to the public. According to the normalized trend, the community seems to exhaust most beneficial approaches in about one month after the release of a model. The exploitation does not stop after that point, but the improvements are slowing down. Mistral 7 billion and Llama 1 were sustained and enhanced for an extended period following their release, with an approximate 15 percent enhancement during this period. This indicates that these models remain valuable and relevant to the community, with ongoing efforts to boost their performance. Furthermore, the excitement and anticipation surrounding the release of new models can drive rapid initial improvements, underscoring the importance of community engagement and the hype cycle in the open source AI ecosystem. 

The data also highlights differences in the trainability of these models compared to, for example, phi models, which are relatively small and likely optimised to their fullest extent by Microsoft, or larger models that may exceed the resources available to most open-source contributors.

\section{Conclusion}

The ongoing debate about the openness of AI models is critically important, as the implications of adopting either extreme—completely closed or fully open—are profound. When addressing decisions of such significance, it is crucial to consider as many factors as possible. This paper aims to contribute to the discussion by offering insights through a data-driven approach. While we strive to maintain objectivity in interpreting the results, some conclusions are clear and are thus emphasised.

Our findings indicate that the open-source community is expanding rapidly, attracting talent from across the world. This influx of contributors plays a vital role in enhancing existing models and ensuring stable development in the future. Furthermore, the open-source community serves as a significant source of new ideas and approaches, which private enterprises can access freely. 

However, the community in a large extend depends on businesses that focus on developing base models and releasing them to the community. In such cases, businesses developing proprietary AI models often face the challenge of safeguarding their intellectual property to maintain a competitive edge. We argue that traditional business models for proprietary software may not fully align with the unique characteristics of AI. A key distinction is the inability of the general public to privately run these models. Consequently, the future of AI development may lean toward a software-as-a-service (SaaS) model. This could create a unique dynamic where the open-source community contributes to model development, while enterprises generate revenue from model usage, benefiting from a continuous flow of ideas and improvements from the open-source ecosystem.

Finally, this research aims to assist policymakers in shaping regulations for the AI industry. Policymakers face a dual responsibility of protecting the public from misuse while safeguarding enterprises and ensuring a fair competitive landscape to promote future innovation.

Future work should focus on assessing whether Hugging Face benchmarks accurately reflect real-world model performance. Additionally, pairing the dataset with scientific literature could enable a chronological analysis of how various methods are introduced. This could provide valuable insights into whether the open-source community primarily contributes engineering solutions or also generates novel ideas that later influence academic research.

\bibliographystyle{IEEEtran}
\bibliography{bibliography}

\EOD

\end{document}